\documentclass[11pt,showpacs]{revtex4}
\def\comment#1{}

\usepackage{graphicx}
\begin{document}
\title{Quantum Regge Calculus of Einstein-Cartan theory }
\author{She-Sheng Xue}
\email{xue@icra.it}
\affiliation{ICRANeT Piazzale della Repubblica 10, 65122 Pescara, Italy
}


\begin{abstract}
We study the Quantum Regge Calculus 
of Einstein-Cartan theory to describe quantum dynamics of Euclidean space-time discretized as a 4-simplices complex.
Tetrad field $e_\mu(x)$ and spin-connection field $\omega_\mu(x)$ are assigned to 
each 1-simplex. Applying the torsion-free Cartan structure equation to each 2-simplex, 
we discuss parallel transports and  
construct a diffeomorphism and 
{\it local} gauge-invariant Einstein-Cartan 
action. Invariant  
holonomies of tetrad and spin-connection fields along large loops 
are also given.  
Quantization is defined by a bounded 
partition function with the measure of $SO(4)$-group valued $\omega_\mu(x)$ fields and Dirac-matrix valued 
$e_\mu(x)$ fields over 4-simplices complex.
\comment{  
In the 2-dimensional case (2-simplices complex), 
we calculate: (i) system's entropy and free-energy, being proportional to its surface; 
(ii) the average of regularized Einstein-Cartan action, 
implying that the Planck length sets the scale for 
the minimal distance between two space-time points. calculations of partition function, entropy and averaged EC 
action in 2-dimensional case.    
}
\end{abstract}
\pacs{04.60.-m,11.10.-z,04.60.Nc,11.15.Ha,05.30.-d}
\maketitle

\noindent
{\it Introduction.}
\hskip0.1cm
Since the Regge Calculus \cite{regge61} was proposed for the discretization of gravity theory in 1961, many progresses have been made in the approach of Quantum Regge Calculus \cite{hammer_book,shamber}
and its variant dynamical triangulations \cite{loll1999}. In particular,
the renormalization group treatment is applied to 
discuss any possible scale dependence of gravity \cite{hammer_book}.
In Lagrangian formalism, gauge-theoretic formulation 
\cite{smolin1979} of quantum gravity using connection variables on a flat hypercubic lattice of the space-time was  
inspired by the success of 
lattice regularization of
non-Abelian gauge theories. 
A locally finite model for gravity has been recently proposed \cite{thooft2008}.
In this Letter, based on the scenario of Quantum Regge Calculus, we present a diffeomorphism and 
{\it local} gauge-invariant invariant regularization and quantization of Euclidean Einstein-Cartan (EC) theory, invariant  
holonomies of tetrad and spin-connection fields $\omega_\mu(x)$ along large loops in 4-simplices complex, and some calculations in 2-dimensional case. 

\vskip0.1cm
\noindent
{\it Euclidean Einstein-Cartan gravity.}
\hskip0.1cm
The basic gravitational variables in the Einstein-Cartan gravity constitute a pair of tetrad and spin-connection fields 
$(e_\mu^a, \omega^{ab}_\mu)$, whose Dirac-matrix values $e_\mu = e_\mu^a\gamma_a$ and $\omega_\mu = \omega^{ab}_\mu\sigma_{ab}$. 
The space-time metric of 4-dimensional Euclidean manifold ${\mathcal M}$ is
$g_{\mu\nu}(x)=e^a_\mu(x)e^b_\nu(x)\delta_{ab}$, where $\delta^{ab}=(+,+,+,+)$.
The diffeomorphism invariance under general coordinate transformations $x\rightarrow x'(x)$ is preserved by all derivatives and $d$-form fields on ${\mathcal M}$ made to be coordinate scalars with the help of tetrad fields $e^a_\mu=\partial\xi^a/\partial x^\mu$. Under the local Lorentz coordinate transformation $\xi^{'a}(x)=[\Lambda(x)]^a_b\xi^b(x)$, 
the {\it local} (w.r.t $\xi$) gauge transformations are: 
\begin{eqnarray}
e'_\mu(\xi)&=&\!{\mathcal V}(\xi)e_\mu(\xi){\mathcal V}^\dagger(\xi),\label{varie}\\
\omega'_\mu(\xi)&=&{\mathcal V}(\xi)\omega_\mu(\xi){\mathcal V}^\dagger(\xi)+
{\mathcal V}(\xi)\partial_\mu{\mathcal V}^\dagger(\xi);
\label{gtran0}
\end{eqnarray}
and fermion field $\psi'(\xi)={\mathcal V}(\xi) \psi(\xi)$, the covariant derivative ${\mathcal D}'_\mu={\mathcal V}(\xi){\mathcal D}_\mu
{\mathcal V}^\dagger(\xi)$, ${\mathcal D}_\mu =\partial_\mu - ig\omega_\mu(\xi)$ 
where $g$ is the gauge coupling, $\partial_\mu= e_\mu^a(\partial/\partial\xi^a)$,
${\mathcal V}(\xi)=\exp i[\theta^{ab}(\xi)\sigma_{ab}]\in SO(4)$, and $\theta^{ab}(\xi)$ is an arbitrary function of $\xi$.
In an $SU(2)$ gauge theory, 
gauge field $A_a(\xi_E)$ can be viewed as a connection 
$\int A_a(\xi_E)d\xi_E^a$ on the global flat manifold. On a locally flat manifold, the spin-connection 
$\omega_\mu dx^\mu =\omega_a(\xi)d\xi^a$, 
where $\omega_a(\xi)=e^\mu_a\omega_\mu$,
one can identify that the spin-connection field 
$\omega_\mu(x)$ or $\omega_a(\xi)$ is the gravity analog of gauge field
and its {\it local} curvature is given by
\begin{equation}
R^{ab}=d\omega^{ab} - g\omega^{ae}\wedge\omega^{b}{}_{e},
\label{rcurvature}
\end{equation}
and $
R^{'ab}={\mathcal V}(\xi)R^{ab}(\xi){\mathcal V}^\dagger(\xi)$ under the transformation (\ref{varie},\ref{gtran0}).
The diffeomorphism and {\it local} gauge-invariant EC action for gravity 
is given by 
the Palatini action $S_P$ 
and Host modification $S_H$ 
\begin{eqnarray}
S_{EC}(e,\omega
)&=&S_P(e,\omega)+S_P(e,\omega)
\label{ec0}\\
S_P(e,\omega)&=&\frac{1}{4\kappa}\int_{\mathcal M} d^4x\det(e)\epsilon_{abcd}e^a\wedge e^b \wedge R^{cd},
\label{host}\\
S_H(e,\omega)&=&\frac{1}{2\kappa\tilde\gamma}\int_{\mathcal M} d^4x\det(e) e_a\wedge e_b \wedge R^{ab}\label{host1}
\end{eqnarray}
where $\kappa\equiv 8\pi G$, the Newton constant $G=1/m_{\rm Planck}^2$, and $\det(e)$ is the Jacobi of mapping $x\rightarrow \xi(x)$.
\comment{
In addition, 
the diffeomorphism invariance under the general coordinate transformation $x\rightarrow x'(x)$ 
is preserved by all fields in Eqs.~(\ref{host}-\ref{art}) 
made to be coordinate scalars by using tetrad fields. The derivatives represents the propagation of fields in coordinate space is related to the connection field 
in local Lorentz frame. spin-connection is the gravity analog of gauge field, how ever it is constructed by the general coordinate derivatives of tetrad field relating to general connection. .....
}
The complex Ashtekar connection \cite{A1986} with 
reality condition and the real Barbero connection \cite{b1995} are linked by a canonical transformation of the connection with a finite complex 
Immirzi parameter $\tilde\gamma\not=0$ \cite{i1997}, which is crucial for {\it Loop Quantum Gravity} \cite{rt1998}.
\comment{, a quantum theory of gravity in 
Hamiltonian formalism, where intrinsic discrete eigenvalues of invariant area and volume operators are obtained
in the diffeomorphism invariant Hilbert space, as results, 
the space-time is discretized with the Planck length and
the black-hole entropy is obtained.
} 

Classical equations can be obtained by the invariance of the EC action (\ref{ec0})
under the transformation (\ref{varie}-\ref{gtran0}),
\begin{equation}
\delta S_{EC}=\frac{\delta S_{EC}}{\delta e_\mu}\delta e_\mu
+\frac{\delta S_{EC}}{\delta \omega_\mu}\delta \omega_\mu
=0,
\label{inv}
\end{equation}
where $\delta e_\mu 
$ and $\delta\omega_\mu$  are infinitesimal variations, which can be expressed in terms of independent Dirac matrix bases $\gamma_5$ and $\gamma_\mu$. 
Therefore, for an arbitrary function $\theta_{ab}$, we have
$\delta S_{EC}/\delta e_\mu=0$ and 
$\delta S_{EC}/\delta \omega_\mu=0$, respectively leading to
Einstein equation and
Cartan's structure equation (torsion-free)
\begin{equation}
de^a -\omega^{ab}\wedge e_{b}
=0.
\label{werelation1}
\end{equation} 

\begin{figure}[ptb]
\includegraphics[width=4.8cm,height=2.5cm]{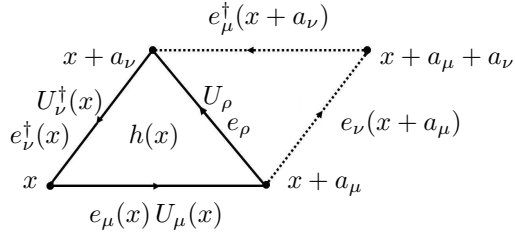}
\put(-100,25){$h(x)$}
\put(-140,8){$x$}
\put(-40,8){$x+a_\mu$}
\put(-115,-5){$e_\mu(x)$}
\put(-90,-5){$U_\mu(x)$}
\put(-5,55){$x+a_\mu+a_\nu$}
\put(-125,55){$x+a_\nu$}
\put(-70,70){$e^\dagger_\mu(x+a_\nu)$}
\put(-145,25){$e^\dagger_\nu(x)$}
\put(-135,38){$U^\dagger_\nu(x)$}
\put(-63,30){$e_\rho$}
\put(-72,40){$U_\rho$}
\put(-20,30){$e_\nu(x+a_\mu)$}
\caption{
Assuming edge spacing $a_{\mu,\nu}(x)$ is 
so small that the geometry of the interior of 4-simplex and its sub-simplex (3- and 2-simplex) is approximately flat, we assign a local Lorentz frame to each 4-simplex. On a local Lorentz manifold $\xi^a(x)$ at a space-time point 
``$x$'', we sketch a closed parallelogram ${\mathcal C}_P(x)$ lying in the 2-simplex $h(x)$. Its edges 
$e_\mu(x)$ and $e^\dagger_\nu(x)=e_\nu(x+a_\nu)$ are two edges of the 2-simplex $h(x)$, and other edges (dashed lines) $e^\dagger_\mu(x+a_\nu)$ and $e_\nu(x+a_\mu)$ 
are parallel transports of $e_\mu(x)$ and $e^\dagger_\nu(x)$ along $\nu$- and $\mu$-directions respectively. 
Each 2-simplex in the 4-simplices complex has a closed parallelogram lying in it.
Group-valued gauge fields $U_\mu(x)$ and $U^\dagger_\nu(x)=U_\nu(x+a_\nu)$ are respectively associated to edges $e_\mu(x)$ and $e^\dagger_\nu(x)$ of the 2-simplex $h(x)$, as indicated. The fields $e_\rho(x+a_\mu)$ and $U_\rho(x+a_\mu)$ are associated to the third edge $(x+a_\mu,x+a_\nu)$ of the 2-simplex $h(x)$. 
}%
\label{pl}%
\end{figure}

\vskip0.1cm
\noindent
{\it Regularized EC action.}
\hskip0.1cm 
The four-dimensional Euclidean manifold ${\mathcal M}$ 
is discretized as an ensemble of ${\mathcal N}_0$ 
space-time points ``${\it x}$'' and ${\mathcal N}_1$ 
links (edges) ``$l_\mu(x)$'' connecting two 
neighboring points, which is a simplicial manifold. The way to construct a simplicial manifold depends also on the assumed topology of the manifold, which gives geometric constrains on the numbers of sub-simplices (${\mathcal N}_0,{\mathcal N}_1,\cdot\cdot\cdot$, see 
Ref.~\cite{loll1999}). 
In this Letter, analogously to the simplicial manifold adopted by Regge Calculus 
we consider a 4-simplices complex, whose 
elementary building block is a 4-simplex (pentachoron). 
The 4-simplex has $5$ vertexes -- 0-simplex (a space-time point ``${\it x}$''), $5$ ``faces'' -- 3-simplex (a tetrahedron), and each 3-simplex has 4 faces -- 2-simplex (a triangle), and each 2-simplex has three faces -- 1-simplex (an edge or a link ``$l_\mu(x)$''). 
Different configurations of 4-simplices complex correspond to variations of relative vertex-positions $\{x\}$, edges ``\{$l_\mu(x)\}$'' and 
``deficit angle'' around each vertex $x$.
These configurations will be described by the configurations of dynamical fields $e_\mu(x)$ and $\omega_\mu(x)$ (its group-valued $U_\mu(x)$) 
in a regularized EC-theory \cite{FT}.   
 
To illustrate how to construct a regularized EC theory describing dynamics of 4-simplices complex, we consider a 
2-simplex (triangle) $h(x)$ (see Fig.~\ref{pl}).
The fundamental tetrad field $e_\mu(x)$ and ``gauge'' field $\omega_\mu(x)$ are assigned to each 1-simplex (edge) of the 4-simplices complex. The values of $e_\mu(x)$-field characterize edge spacings $a_\mu(x)\equiv |l_\mu(x)|$, where $l_\mu(x)=ae_\mu(x)$ and the Planck length $a=(8\pi G)^{1/2}$. The fundamental area operator $S^{\rm h}_{\mu\nu}\equiv l_\mu(x)\wedge l_\nu(x)/2$, where $\mu\not=\nu$ indicates edges of the 2-simplex.
The 2-simplex area $S_{\rm h}(x)=|S^{\rm h}_{\mu\nu}(x)|$. 

The Cartan equation (\ref{werelation1}) is actually an equation for infinitesimal parallel transports of $e_\nu(x)$ fields. Applying this equation to the 2-simplex $h(x)$, as shown in Fig.~{\ref{pl}},  we show that $e_\nu(x)$ [$e_\mu(x)$] undergoes its 
parallel transport to $e_\nu(x+a_\mu)$ [$e_\mu(x+a_\nu)$] 
along the $\mu$ [$\nu$]-direction for an edge spacing $a_\mu(x)$ [$a_\nu(x)$], following 
the discretized Cartan equation
\begin{eqnarray}
e^a_\nu(x+a_\mu)-e^a_\nu(x) -a_\mu\omega_\mu^{ab}(x)\wedge e_{\nu b}(x)&=&0,
\label{wel1}
\end{eqnarray}
and $\mu\leftrightarrow\nu$.
The parallel transports $e^a_\nu(x+a_\mu)$ and $e^a_\mu(x+a_\nu)$ are neither independent fields, nor assigned to any edges of the 4-simplices complex. 
They are related 
to $e_\mu(x)$ and $\omega_\mu(x)$ fields assigned to edges of the 2-simplex $h(x)$ by the Cartan equation (\ref{wel1}). Because of 
torsion-free, $e_\mu(x),e_\nu(x)$ and their parallel transports $e_\mu(x+a_\nu),e_\nu(x+a_\mu)$ form a {\it closed} parallelogram ${\mathcal C}_P(x)$ (Fig.~\ref{pl}). Otherwise this would means the curved space-time could not be approximated locally by a flat space-time \cite{hrbook}. 
\comment{
Thus, for each 2-simplex, there is a closed parallelogram, whose two edges lying in the  2-simplex and other two edges of parallel transports not lying in any 2-simplex.  
}
We define $\omega_\mu(x+a_\nu)$ and 
$\omega_\nu(x+a_\mu)$ by using the discretized equation for curvature (\ref{rcurvature}),
\begin{eqnarray}
\omega^{ab}_\nu(x\!+\!a_\mu) 
\!-\! \omega^{ab}_\nu(x) \!-\! a_\mu\omega^{ae}_\mu(x)\wedge \omega^b_{e\nu}(x)\!\! &=& \!\! a_\mu R^{ab}_{\mu\nu}(x),\label{ww1}
\end{eqnarray}
and $\mu\leftrightarrow\nu$.  For zero curvature case, analogously to (\ref{wel1}),
parallel transports $\bar\omega^{ab}_\nu(x+a_\mu)$ [$\bar\omega^{ab}_\mu(x+a_\nu)$] can be defined as 
\begin{eqnarray}
\bar\omega^{ab}_\nu(x+a_\mu) 
- \omega^{ab}_\nu(x) - a_\mu\omega^{ae}_\mu(x)\wedge \omega^b_{e\nu}(x) &=&0,
\label{ww1p}
\end{eqnarray}
and $\mu\leftrightarrow\nu$. 
The difference (``deficit angle'') between $\omega^{ab}_\nu(x+a_\mu)$ and $\bar\omega^{ab}_\nu(x+a_\mu)$ is the curvature $a_\mu R^{ab}_{\mu\nu}(x)$. 

Instead of $\omega_\mu(x)$ field, we assign a group-valued field $U_\mu(x)$ to each 1-simplex of 4-simplices complex. For example, at edges $(x,\mu)$ and $(x,\nu)$ of the 2-simplex $h(x)$ ($\mu\not=\nu$ see Fig.~\ref{pl}), 
we define $SO(4)$ group-valued spin-connection fields,
\begin{eqnarray}
U_\mu(x)= e^{iga\omega_\mu(x)},\quad U_\nu(x)= e^{iga\omega_\nu(x)},
\label{link0}
\end{eqnarray}
which 
take value of fundamental representation of the compact group $SO(4)$, and
their local gauge transformations,
\begin{eqnarray}
U_\mu(x)&\rightarrow &{\mathcal V}(x)U_\mu(x){\mathcal V}^\dagger(x+a_\mu),
\label{gtranu}
\end{eqnarray}
and $\mu\leftrightarrow\nu$ in accordance with (\ref{gtran0}).
Actually, these group-valued fields (\ref{link0}) can be viewed as unitary operators for finite parallel transportations. Eq.~(\ref{wel1}) can be 
generalized to
\begin{eqnarray}
e_\nu(x+a_\mu)&=& U_\mu(x)e_\nu(x)U^\dagger_\mu(x),
\label{up1}
\end{eqnarray}
and $\mu\leftrightarrow\nu$. While, corresponding to (\ref{ww1}) for the field $\omega_\nu(x+a_\mu)$, we define  
\begin{eqnarray}
 U_\nu(x+a_\mu)&\equiv &U_\mu(x)U_\nu(x)U^\dagger_\mu(x),\label{up11}\\
 U_\nu(x+a_\mu) &\equiv& e^{iga\omega_\nu(x+a_\mu)},
\label{link00}\\
U_{\mu\nu}(x)&\equiv &U_\mu(x) U_\nu(x)\equiv U_\nu(x+a_\mu)U_\mu(x),
\label{p00}
\end{eqnarray}
and $\mu\leftrightarrow\nu$. Eq.~(\ref{p00}) characterizes relative angles $\theta_{\mu\nu}(x)$ between two neighboring edges $e_\mu(x)$ and $e_\nu(x)$ 
(see Fig.~\ref{pl}). 
In the {\it naive continuum limit}: 
$ag\omega_\mu \ll 1$ (small coupling or weak-field), indicating that the wavelengths of 
weak and slow-varying fields $\omega_\mu(x)$  
are much larger than the edge spacing $a_{\mu,\nu}$, 
we have
\begin{eqnarray}
U_{\mu\nu}(x)
&=&\exp\Big\{ig[a\omega_\nu(x)+a\omega_\mu(x)]
+iga^2\partial_\mu \omega_\nu(x)\nonumber\\
&-&\frac{1}{2}(ga)^2\left[\omega_\nu(x),\omega_\mu(x)\right]+{\mathcal O}(a^3)\Big\},
\label{uu1}
\end{eqnarray}
where ${\mathcal O}(a^3)$ indicates 
high-order powers of $ag\omega_\mu$. 

Using the tetrad fields $e_\mu(x)$ to construct coordinate and Lorentz scalars so as to obtain a regularized EC action preserving the diffeomorphism 
and {\it local} gauge-invariance, we define 
the smallest holonomy along closed triangle path of 2-simplex:
\begin{eqnarray}
X_{h} (v,U)&=& 
{\rm tr}\left[v_{\nu\mu}(x)U_{\mu}(x)v_{\mu\rho}(x+a_\mu)U_{\rho}(x+a_\mu)
v_{\rho\nu}(x+a_\nu)U_{\nu}(x+a_\nu)\right],
\label{xs}
\end{eqnarray}
whose orientation is anti-clock-like, and $X^\dagger_{h}(e,U)$ 
is clock-like (see Fig.~\ref{pl}). 
We have following two possibilities for the vertex-field $v_{\nu\mu}(x)$. The first
$v_{\mu\nu}(x)  =  e_{\mu\nu}(x)\gamma_5$:   
\begin{eqnarray}
{\mathcal A}_P(e,U)&=&\frac{1}{8g^2}\sum_{h}
\left\{X_{h} (v,U)+{\rm h.c.}\right\},
\label{pact}\\
e_{\mu\nu}(x)& \equiv & 
(e^a\wedge e^b)\sigma_{ab},
\label{dirace}
\end{eqnarray}
where 
$\sum_{h}$ is the sum over all 2-simplices $h(x)$. 
In the limit: $ag\omega_\mu \ll 1$, Eq.~(\ref{pact}) becomes
\begin{eqnarray}
{\mathcal A}_P(e,U_\mu)
&= &\frac{1}{a^2} \sum_{h} S_{\rm h}^2(x) \epsilon_{cdab}\, e^c\wedge e^d \wedge R^{ab}+{\mathcal O}(a^4).
\label{pa2}
\end{eqnarray}
We define a 4-d volume element $V(x)=\sum_{h(x)} S_{\rm h}^2(x)$ around the vertex $x$. The interior of 4-simplex is approximately flat, leading to
\begin{eqnarray}
\sum_{x}V(x) \Rightarrow \int d^4\xi(x)
=\int d^4x{\rm det}[e(x)],
\label{vint}
\end{eqnarray}
and Eq.~(\ref{pa2}) approaches to $S_P(e,\omega)$ (\ref{host}) with an effective Newton constant $G_{\rm eff}=g G/4$.
The second $v_{\mu\nu}(x)  =  e_{\mu\nu}(x)$:
\begin{eqnarray}
{\mathcal A}_H(e,U_\mu)&=&\frac{1}{8g^2\gamma}\sum_{h}
\left[X_h(v,U)+{\rm h.c.}\right],\label{hact}
\label{diract}
\end{eqnarray}
where the real parameter $\gamma=i\tilde\gamma$.
Analogously, in the limit: $ag\omega_\mu \ll 1$, Eq.~(\ref{hact}) approaches to $S_H(e,\omega)$ (\ref{host1}),
\begin{eqnarray}
{\mathcal A}_H(e,U_\mu)
&= & \frac{1}{2\kappa \tilde\gamma}\int d^4x{\rm det}[e(x)] 
 e_a\wedge e_b \wedge R^{ab}+{\mathcal O}(a^4).
\label{hpa2}
\end{eqnarray}
Under the gauge transformation (\ref{varie}), 
\begin{eqnarray}
v_{\mu\nu}(x)&\rightarrow &{\mathcal V}(x)v_{\mu\nu}(x){\mathcal V}^\dagger(x).
\label{gtranl}
\end{eqnarray}
The diffeomorphism 
and {\it local} gauge-invariant regularized EC action is then given by
\begin{eqnarray}
{\mathcal A}_{EC}={\mathcal A}_{P}+{\mathcal A}_{H}.
\label{ecp}
\end{eqnarray}
\comment{
$a=\pi/\Lambda_{\rm cutoff}$, the momentum cutoff $\Lambda_{\rm cutoff}=m_p(\pi/8)^{1/2}$
}

Considering the following diffeomorphism and {\it local} gauge-invariant holonomies along a large loop ${\mathcal C}$ on the Euclidean manifold ${\mathcal M}$ 
\begin{eqnarray}
X_{\mathcal C}(v,\omega)&=&
{\mathcal P}_C{\rm Tr}\exp\left\{ ig\oint_{\mathcal C}v_{\mu\nu}(x)
\omega^\mu(x) dx^\nu\right\},
\label{pa0s}
\end{eqnarray} 
where ${\mathcal P}_C$ is the path-ordering and ``${\rm Tr}$'' denotes 
the trace over spinor space, we attempt to regularize these holonomies
on the 4-simplices complex. 
Suppose that an orientating closed path ${\mathcal C}$ 
passes space-time points $x_1,x_2,x_3,\cdot\cdot\cdot, x_N=x_1$ and edges connecting between neighboring points in the 4-simplices complex. 
At each point $x_i$
two tetrad fields $e_\mu(x_i)$ and $e_{\mu'}(x_i)$ $(\mu\not=\mu')$ respectively 
orientating path incoming to ($i-1\rightarrow i$) and outgoing from 
($i\rightarrow i+1$) the point $x_i$, we have the 
vertex-field $v_{\mu\mu'}(x_i)$ defined by Eqs.~(\ref{dirace},\ref{diract}).  
Link fields $U_\mu(x_i)$ are defined on edges lying in the loop ${\mathcal C}$, recalling the relationship 
$U_\mu(x_i)=U_{-\mu}(x_{i+1})=U^\dagger_{\mu}(x_{i+1})$, 
we can write the regularization of the holonomies (\ref{pa0s}) as follows,
\begin{eqnarray}
X_{\mathcal C}(v,U)&\!\!=\!\!&{\mathcal P}_C{\rm Tr} \Big[v_{\mu\mu'}(x_1)U_{\mu'}(x_1)v_{\mu'\nu}(x_2)U_{\nu}(x_2)\nonumber\\
&\cdot\cdot\cdot &
v_{\rho\rho'}(x_i)U_{\rho'}(x_i)v_{\rho'\sigma}(x_{i+1})\nonumber\\
&\cdot\cdot\cdot &v_{\lambda\mu}(x_{N-1})U^\dagger_{\mu}(x_{N-1})\Big],
\label{rpa0s}
\end{eqnarray} 
preserving diffeomorphism and {\it local} gauge-invariances. Eq.~(\ref{rpa0s}) is  consistent with Eq.~(\ref{xs}). 

\vskip0.1cm
\noindent
{\it Euclidean partition function.}
\hskip0.1cm
The partition function $Z_{EC}$ and effective action 
${\mathcal A}^{\rm eff}_{EC}$ are
\begin{eqnarray}
Z_{EC}=\exp-{\mathcal A}^{\rm eff}_{EC}=\int {\mathcal D}e{\mathcal D}U
\exp -{\mathcal A}_{EC},
\label{par}
\end{eqnarray}
with the diffeomorphism and {\it local} gauge-invariant measure  
\begin{eqnarray}
\int \!\!{\mathcal D}e{\mathcal D}U 
\!\equiv\!\! \prod_{x,\mu}\!\!\int\!\! de_\mu(x)dU_\mu(x)
\label{mea1}
\end{eqnarray}
where $\prod_{x,\mu}$ indicates the product of overall edges, $dU_\mu(x)$ is the Haar measure of compact gauge group $SO(4)$ or $SU(2)$, and $de_\mu(x)$ is the measure of Dirac-matrix valued field $e_\mu(x)=\sum_a e_\mu^a(x)\gamma_a$, determined by the functional measure $de_\mu^a(x)$ of the bosonic field $e_\mu^a(x)$. It should be mentioned that the measure (\ref{mea1}) is just a lattice form of the standard DeWitt functional measure \cite{dewitt67} over the continuum degrees, with the integral of the spin-connection field $\omega_\mu(x)$ replaced by the Haar integral over the $U_\mu(x)$'s, analytical integration or numerical simulations runs overall configuration space of continuum degrees and no gauge fixing is needed.
\comment{   
Note that the measure ${\mathcal D}U_\mu(x)$ 
includes all link fields lying in both edges ($e_\mu,e_\nu$) of 2-simplices and their parallel transports ($e_\mu,e_\nu$), as shown in Fig.~\ref{pl}.
\begin{eqnarray} 
[e^a_\mu(x),e^b_\nu(x')]=\delta_{\mu\nu}(x)\delta^{ab}\delta(x-x'),
\label{qe0}
\end{eqnarray}
and equivalently
\begin{eqnarray} 
\{e_\mu(x),e^\dagger_\nu(x')\}=\delta_{\mu\nu}(x)\delta(x-x').
\label{qe}
\end{eqnarray}
}
In this path-integral quantization formalism, values of the partition function (\ref{par}) presents 
all dynamical configurations of 4-simplices complex, described by field configurations $e_\mu(x)$ and ${U_\mu(x)}$ in the weight $\exp -{\mathcal A}_{EC}$. The vacuum expectational 
values (v.e.v.) of diffeomorphism and {\it local} gauge-invariant quantities, for instance holonomies (\ref{rpa0s}), are given by
\begin{eqnarray}
\langle X_{\mathcal C}(e,U)\rangle=\frac{1}{Z_{EC}}\int {\mathcal D}e{\mathcal D}U
\Big[X_{\mathcal C}(e,U)\Big]\exp -{\mathcal A}_{EC}\,.
\label{eve}
\end{eqnarray} 
In the action (\ref{pact},\ref{hact}), $X_h(v,U)$ (\ref{xs}) contains the quadric term of $e_\mu(x)$-field associated to each edge $(x,\mu)$, 
the partition function $Z_{EC}$ (\ref{par}) and v.e.v.~(\ref{eve}) are converge. 
\comment{
and we have the following formula:
\begin{eqnarray}
\int {\mathcal D}e 
e^{ -e_l\Delta^{lk}(U)e^\dagger_k}&=& \det[\Delta(U)],
\label{eint1}\\
\int {\mathcal D}e (e_i e^\dagger_j)e^{ -e_l\Delta^{lk}(U)e^\dagger_k}
&=& \Delta_{ij}(U),
\label{eint2}\\
\int {\mathcal D}e [e_i \Lambda^{ij}(U) e^\dagger_j]e^{ -e_l\Delta^{lk}(U)e^\dagger_k}
&=& \text{Tr}[\Lambda(U)\Delta(U)],
\label{eint3}
\end{eqnarray}
where $\Delta(U)$ and $\Lambda(U)$ are operators in terms of links fields $\{U_\mu(x)\}$.
Applying Eq.~(\ref{eint1}) to the partition function (\ref{par}), 
we integrate over tetrad fields $e_\mu(x)$ and formally obtain, 
\begin{eqnarray}
Z_{EC}=\int {\mathcal D}U \det\Big[ \frac{1}{8g}\gamma_5 U_{\mu\nu}\frac{i}{2}+ \frac{1}{8g\gamma}U_{\mu\nu} +{\rm h.c.} \Big ].
\label{par1}
\end{eqnarray}
}  

Analogously to Eq.~(\ref{inv}), 
the {\it local} gauge-invariance of the partition function (\ref{par}) ($\delta Z_{EC}=0$) leads to
\begin{equation}
\langle \frac{\delta{\mathcal A}_{EC}}{\delta e_\mu}\delta e_\mu
+U_\mu\frac{\delta {\mathcal A}_{EC}}{\delta U_\mu}
+{\rm h.c.}\rangle =0,
\label{inv1}
\end{equation}
which becomes ``averaged'' Einstein equation $\langle{\delta \mathcal A}_{EC}/\delta e_\mu\rangle +{\rm h.c.}=0$, and
\begin{equation}
\langle U_\mu\frac{\delta {\mathcal A}_{EC}}{\delta U_\mu}
- U^{\dagger }_\mu\frac{\delta {\mathcal A}_{EC}}{\delta U^{\dagger}_\mu}\rangle =0.
\label{inv2}
\end{equation}
Eq.~(\ref{inv2}) is ``averaged'' torsion-free Cartan equation (\ref{werelation1}), which actually
shows the impossibility of spontaneous breaking of {\it local} gauge symmetry. This should not be surprised, since the torsion-free (\ref{werelation1}) is a necessary condition to 
have a {\it local} Lorentz frame, therefore a {\it local} gauge-invariance.

The {\it local} gauge-invariance of (\ref{eve}) ($\delta \langle X\rangle =0 $) leads to dynamical equations for holonomies (\ref{rpa0s}), which can be formally written as
\begin{equation}
\langle \frac{\delta X}{\delta e_\mu}\delta e_\mu +X\frac{\delta{\mathcal A}_{EC}}{\delta e_\mu}\delta e_\mu
+X+XU_\mu\frac{\delta {\mathcal A}_{EC}}{\delta U_\mu}
+{\rm h.c.}\rangle =0,
\label{sinv1}
\end{equation}
leading to $\langle \delta X/\delta e_\mu+X{\delta \mathcal A}_{EC}/\delta e_\mu\rangle +{\rm h.c.}=0$, and
\begin{equation}
\langle X\rangle + \langle X\Big( 
U_\mu\frac{\delta {\mathcal A}_{EC}}{\delta U_\mu}
- U^{\dagger }_\mu\frac{\delta {\mathcal A}_{EC}}{\delta U^{\dagger}_\mu}\Big)\rangle =0.
\label{sinv2}
\end{equation}
Eq.~(\ref{sinv2}) has the same form as the Schwinger-Dyson equation for Wilson loops in lattice gauge theories. 

The regularized EC theory (\ref{ecp}) can be separated into left- and right-handed parts 
by replacing  $U_\mu(x)=U^L_\mu(x)\otimes U^R_\mu(x)$, 
where $U^{L,R}_\mu(x)\in SU_{L,R}(2)$. In addition,
we can generalize the link field $U_\mu(x)$
to be all irreducible representations $U^j_\mu(x)$ of the gauge group $SO(4)$. 
The regularized EC action (\ref{ecp}) should be a sum over all representations $j\equiv j_{L,R}=1/2,3/2,\cdot\cdot\cdot$,
\begin{equation}
{\mathcal A}_{EC}=\sum_j\left[
{\mathcal A}^j_P(e_\mu,U^j_\mu)+{\mathcal A}^j_H(e_\mu,U^j_\mu)
\right],
\label{allj}
\end{equation}
and the measure (\ref{mea1}) should include all representations of gauge group.

\vskip0.1cm
\noindent
{\it Some calculations in 2-dimensional case. 
}
\hskip0.1cm
We consider a 2-simplices complex, i.e., random simplicial surface, 
whose elementary building block is a triangle $h(x)$ (see Fig.~\ref{pl}).  
In this case, 
{\it local} gauge transformations (\ref{gtranu},\ref{gtranl}) can be made so that all fields $v_{\mu\rho}(x+a_\mu)U_{\rho}(x+a_\mu)
v_{\rho\nu}(x+a_\nu)=1$ in Eq.~(\ref{xs}), as if we choose a particular gauge. 
The partition function (\ref{par}) can be calculated by integrating over $e_\mu(x)$- and $U_\mu(x)$-fields, using
the Cayley-Hamilton formula for a determinant \cite{iz4-86} and the properties of invariant Haar measure: 
$\int dU^j_\mu(x)=1$, $\int dU^j_\mu(x)U^j_\mu(x)=0$ and 
\begin{eqnarray}
\int dU^j_\mu(x) U^{ab}_\mu(x)U^{\dagger cd}_\nu(x')=\frac{1}{d_j}\delta_{\mu\nu}\delta^{ac}\delta^{bd}\delta(x-x'),
\label{mea2}
\end{eqnarray}
where $d_j=n_{j_L}n_{j_R}$ ( $n_{j_L,j_R}=2j_{L,R}+1$), the dimension of irreducible representations $j=(j_L,j_R)$ of $SU_L(2)\otimes SU_R(2)$.
\comment{
We calculate Eq.~(\ref{par}) for all representations $j$ 
\begin{eqnarray}
Z_{EC}=\prod\Big[ \frac{i}{2d_jg^2}\gamma_5 + \frac{2}{2d_jg^2\gamma} \Big ].
\label{par2}
\end{eqnarray}
}
We obtain the entropy ${\mathcal S}=\ln Z_{EC}$ 
\begin{eqnarray}
{\mathcal S}= \sum
\text{Tr}\Big[ \gamma_5 \frac{i}{2d_jg^2}+ \frac{2}{2d_jg^2\gamma}  \Big ]=\sum_j\frac{4}{d_jg^2\gamma a^2}S_{\rm surf},
\label{entropy}
\end{eqnarray}
where $\sum$ is the sum over all 2-simplices, degrees of freedom of gauge group representations and Dirac spinors. The 2-dimensional surface
\begin{eqnarray}
S_{\rm surf}= \sum_{h} S_{\rm h}(x)= N_{\rm h} P_a,
\quad P_a=\frac{1}{N_{\rm h}}\sum_{h} S_{\rm h}(x)
\label{surface}
\end{eqnarray} 
where 
$N_{\rm h}$ is the total number of 2-simplices and 
$P_a$ averaged area of 2-simplices. 
The free energy ${\mathcal F}=-\frac{1}{\beta}\ln Z_{EC}$,
\comment{
\begin{eqnarray}
{\mathcal F}=-\frac{1}{\beta}\ln Z_{EC}=-\sum_j\frac{4}{d_j\gamma}S_{\rm surf},
\label{free}
\end{eqnarray}
}
where the inverse ``temperature'' $\beta=1/g^2$, see Eqs.~(\ref{pact},\ref{hact}). 
Selecting fundamental representation 
$d_j=4$, we obtain ${\mathcal S}=S_{\rm surf}/(g^2\gamma a^2)$ 
and ${\mathcal F}=-S_{\rm surf}/(\gamma a^2)$ 
. 

In the same way, we calculate the average of 
regularized EC action ${\mathcal A}_{EC}$ (\ref{allj}), 
\comment{
a single 2-complex action
\begin{equation}
{\mathcal A}^j_{EC}[e_\mu(x),U^j_\mu(x)]=\frac{1}{8g^2}
{\rm tr}\left\{e_{\mu\nu}(x)\gamma_5U^p_{\mu\nu}(x)+\frac{1}{\gamma}\tau_{\mu\nu}(x)U^p_{\mu\nu}(x)
+{\rm h.c.}\right\},
\label{singlea}
\end{equation}
which is
the regularized EC action ${\mathcal A}$ (\ref{allj})
at a single 2-simplex $h(x)$,
i.e., Eqs.~(\ref{pact},\ref{hact}) without the sum $\sum_{x,\mu\nu}=\sum_{h(x)}$. 
}
\comment{
We integrate over tetrad fields $e_\mu(x)$ and obtain,  
\begin{eqnarray}
\langle{\mathcal A}^j_{EC}[e_\mu,U^j_\mu]\rangle &=&\left(\frac{1}{8g^2}\right)^2\frac{1}{Z_{EC}}\int {\mathcal D}U\sum_{h}\cdot\label{eve1}\\
&\cdot &
{\rm tr}\left\{\gamma_5U_{\mu\nu}(x)\left(\frac{i}{2}\right)+\frac{1}{\gamma}U_{\mu\nu}(x)
+{\rm h.c.}\right\}^2,\nonumber
\end{eqnarray}
where $Z_{EC}$ is given by Eq.~(\ref{par}).
In the strong coupling (field) limit $g\gg 1$ or  
$ga\omega_\mu \sim {\mathcal O}(1)$, implying that $\omega_\mu$ field's 
wavelength is comparable to the Planck length $a$,  
we expand $Z_{EC}$ in powers of $1/g$ and use Eq.~(\ref{mea2}) to compute the average (\ref{eve1}). 
As a result,
the leading term is given by 
} 
\begin{equation}
\langle {\mathcal A}^j_{EC}[e_\mu,U^j_\mu] \rangle \simeq \frac{1}{d_j}
\left(\frac{1}{8g^2}\right)^2\left(1+\frac{4}{\gamma^2}\right)N_{\rm h},
\label{aqj}
\end{equation}
in the strong coupling (field) limit $g\gg 1$ or  
$ga\omega_\mu \sim {\mathcal O}(1)$, which implies that $\omega_\mu$ field's 
wavelength is comparable to the Planck length $a$,
The average (\ref{aqj}) of regularized EC action has discrete values 
corresponding to the fundamental state $d_j=4$ 
and excitation states $d_j=16$.
\comment{ 
The average of total regularized EC action 
\begin{equation}
\langle {\mathcal A}^j_{EC}[e_\mu,U^j_\mu] \rangle=\sum_{h(x)}\langle {\mathcal A}^j_{EC}[e_\mu(x),U^j_\mu(x)] \rangle \simeq \frac{1}{2d_jg^2\gamma }N_{\rm h},
\label{taqj}
\end{equation}
where $N_{\rm h}$ is the total number of 2-simplices.
}

Using the convexity inequality $\langle e^{-{\mathcal A}^j_{EC}}\rangle \ge e^{-\langle {\mathcal A}^j_{EC}\rangle}$, 
we have  
\begin{equation}
\langle {\mathcal A}^j_{EC}[e_\mu,U^j_\mu] \rangle \le 
\ln Z^j_{EC}(2/g^2)-\ln Z^j_{EC}(1/g^2). 
\label{minia}
\end{equation}
Using Eqs.~(\ref{entropy},\ref{surface}),
we obtain 
\begin{equation}
\frac{1}{d_j}
\left(\frac{1}{8g^2}\right)^2\left(1+\frac{4}{\gamma^2}\right)N_{\rm h}  \le 
\frac{4}{d_jg^2\gamma a^2}S_{\rm surf}, 
\label{minia1}
\end{equation}
and averaged area of a 2-simplex
\begin{equation}
P_a\ge \frac{\pi}{32 g^2}\left(1+\frac{4}{\gamma^2}\right)\frac{8\pi}{m_{\rm Planck}^2}, 
\label{minia2}
\end{equation}  
implying that the
Planck length is 
minimal separation between two space-time points  \cite{PL}.

\comment{
\begin{equation}
P_a\ge\pi/m_p^2.
\label{miniax}
\end{equation}
Using Eq.~(\ref{aqj}), we show the Planck area $P_a$ has to be larger than $\pi/m_p^2$.
}  

\vskip0.1cm
\noindent
{\it Some remarks.}
\hskip0.1cm
\comment{The quantum dynamics of 4-simplices complex (space time) is described by quantum fields $e_\mu(x)$ and $\omega_\mu(x)$ of regularized and quantized EC theory (\ref{ecp}-\ref{eve}). 
4-simplex, an elementary building block of 4-simplices complex, has the size of order of the Planck length, which is
probed by short wavelengths of 
quantum fluctuations of fields $e_\mu,\omega_\mu$ in strong gauge couplings $g$.  
The genuine violation of the diffeomorphism invariance at the size 
of a 4-simplex is negligible, when we consider
large scales probed by long wavelengths of fields. 
}
\comment{   
We have to point out that the regularization action (\ref{ecp}) is not unique, it can possibly contain non-local high-dimensional ($d>6$) operators of tetrad and link fields, permitted by diffeomorphism and {\it local}  gauge-invariances. 
}
Although the regularized EC action (\ref{ecp}) approaches to the EC action (\ref{ec0}) in the ``{\it naive continuous limit}'' $ag\omega_\mu\ll 1$, the regularized EC theory is physically sensible, provided it has a non-trivial continuum limit. 
It is crucial, on the basis of non-perturbative methods and 
renormalization group invariance, to find: (1) the scaling invariant regimes (ultraviolet fix points) $g_c$, where phase transition takes place and 
physical correlation length $\xi$ is much larger than the Planck length $a$; (2)
$\beta$-function $\beta(g)$ and renormalization-group invariant equation 
$\xi = {\rm const.}\, a\, \exp \int^g dg' /\beta(g')$; (3) all relevant and renormalizable operators (one-particle irreducible (1PI) functions) with effective dimension-4 in these regimes to obtain effective low-energy theories. 
One may add by hand the cosmological $\Lambda$-term $\frac{\lambda}{4\cdot 4!}\epsilon^{\mu\nu\rho\sigma}\sum _x{\rm tr}[e_\mu e_\nu e_\rho e_\sigma]+{\rm h.c.}$, where $\lambda =\Lambda a^2$, into the regularized EC action (\ref{ecp}).
However, 1PI functions ${\mathcal A}^{\rm eff}_{EC}$ (\ref{par}) 
effectively contain this dimensional 
operator, which is related to the truncated Green function 
$\langle {\mathcal A}_{EC}{\mathcal A}_{EC}\rangle$. It is then a question 
what is the scaling property of this operator in terms of $\xi^{-2}$, 
where inverse correlation length $\xi^{-1}$ gives the mass scale of 
low-energy excitations of the theory.     

One can consider the following regularized fermion action, 
\begin{eqnarray}
{\mathcal A}_F(e_\mu,U_\mu,\psi)
&\!\!=\!\!&\frac{1}{2}\sum_{x\mu}\Big[\bar\psi(x) 
e^\mu(x) U_\mu(x)\psi(x+a_\mu)\nonumber\\
&\! \!-\!\!&\bar\psi(x+a_\mu)  U^\dagger_\mu(x)e^\mu(x)\psi(x)\Big],
\label{plart}
\end{eqnarray}
where fermion fields $\psi(x)$ and $\psi(x+a_\mu)$ are defined at two neighboring
points (vertexes) of 4-simplices complex, fields $U_\mu(x)$ and $e_{\mu}(x)$ are added to preserve {\it local} gauge and diffeomorphism invariances, and $\sum_{x\mu}$ is the sum over all edges (1-simplices) of 4-simplices complex.
This bilinear fermion action (\ref{plart}) introduces a non-vanishing torsion field \cite{kleinert,s2001}. We need to study whether the regularized 
EC action (\ref{ecp}) with fermion action (\ref{plart}) can be effectively written in form of a torsion-free part and four fermion 
interactions, as the EC theory in continuum.  
In addition, the bilinear fermion action (\ref{plart}) 
has the problem of either fermion doubling or chiral (parity) 
gauge symmetry breaking, due to the No-Go theorem \cite{nn1981}. Resultant four fermion interactions can possibly be resolution to this 
problem \cite{ep1986,xue1997}. 
Acknowledgment: Author thanks to anonymous referee for his/her comments, 
to H.~Kleinert and J.~ Maldacena for discussions on invariant holonomies in gauge theories.
Author is grateful to H.~W.~Hamber and R.~M.~Williams for discussions on renormalization group invariance and the properties of Dirac-matrix valued tetrad fields.

\comment{
The detailed calculations will be presented in a lengthy article.
It is worthwhile to mention the important issue that bilinear fermion term (\ref{plart}) has the problem of either fermion doubling or chiral (parity) 
gauge symmetry breaking, due to the No-Go theorem \cite{nn1981}. Quasilinear fermionic operators can possibly be resolution to this 
problem \cite{ep1986}, and the four-fermion interactions 
(\ref{4fl1}) was studied in Ref.~\cite{xue1997}.
We have to point out that the regularization action (\ref{ecp}) is not unique, it can possibly contain non-local high-dimensional ($d>6$) operators of tetrad, link and fermion fields, permitted by diffeomorphism and {\it local}  gauge-invariances. Beside, non-local holonomies  (\ref{pa0s}) needs to be defined for a genuine non-perturbative formulation of quantum gravity. The detailed calculations will be presented in a lengthy article.   
}


\begin{thebibliography}{99}
\bibitem{FT} 
We are not clear now how to relate configurations of fields $e_\mu(x)$ and $\omega_\mu(x)$ to topological constrained configurations of 4-simplices complex
in dynamical triangulations.
\end{thebibliography}
\end{document}